\documentclass[a4paper, amsfonts, amssymb, amsmath, reprint, showkeys, nofootinbib, twoside, superscriptaddress]{revtex4-2}
\usepackage[english]{babel}
\usepackage[utf8]{inputenc}
\usepackage{dsfont}
\usepackage{xcolor}
\usepackage[acronym]{glossaries}
\usepackage[colorinlistoftodos, color=green!40, prependcaption]{todonotes}
\usepackage{amsthm}
\usepackage{mathtools}
\usepackage{physics}
\usepackage{xcolor}
\usepackage{graphicx}
\usepackage[left=23mm,right=13mm,top=35mm,columnsep=15pt]{geometry} 
\usepackage{adjustbox}
\usepackage{placeins}
\usepackage[T1]{fontenc}
\usepackage{lipsum}
\usepackage{csquotes}

\newacronym{LVC}{LVC}{Linear Vibronic Coupling}
\newacronym{LHC}{LHC}{light harvesting complex}
\newacronym{CPTP}{CP-TP}{completely positive trace preserving}

\renewcommand{\tr}[1]{\textrm{Tr}\lbrace #1 \rbrace}
\renewcommand{\ket}[1]{\lvert #1 \rangle}
\renewcommand{\bra}[1]{\langle #1 \rvert}

\begin{document}
\title{Computational approaches to efficient generation of the stationary state for
	incoherent light excitation}

\author{Ignacio Loaiza}
    \email[email: ]{ignacio.loaiza@mail.utoronto.ca}% Your name
    \affiliation{Chemical Physics Theory Group, Department of Chemistry, University of Toronto, Toronto, Ontario, M5S 3H6, Canada.}
    \affiliation{Department of Physical and Environmental Sciences, University of Toronto Scarborough, Toronto, Ontario, M1C 1A4, Canada.}
\author{Artur F. Izmaylov}
    \email[email: ]{artur.izmaylov@utoronto.ca}%
    \affiliation{Chemical Physics Theory Group, Department of Chemistry, University of Toronto, Toronto, Ontario, M5S 3H6, Canada.}
    \affiliation{Department of Physical and Environmental Sciences, University of Toronto Scarborough, Toronto, Ontario, M1C 1A4, Canada.}
\author{Paul Brumer}
    \email[email: ]{paul.brumer@utoronto.ca}%
    \affiliation{Chemical Physics Theory Group, Department of Chemistry, University of Toronto, Toronto, Ontario, M5S 3H6, Canada.}

\date{\today} % Leave empty to omit a date

\begin{abstract}
	Light harvesting processes are often computationally studied from a time-dependent viewpoint, in line with ultrafast coherent spectroscopy experiments. Yet, natural processes take place in the presence of incoherent light, which induces a stationary state. Such stationary states can be described using the eigenbasis of the molecular Hamiltonian, but for realistic systems a full diagonalization is prohibitively expensive. We propose three efficient computational approaches to obtaining the stationary state that circumvent system Hamiltonian diagonalization. The connection between the incoherent perturbations, decoherence, and Kraus operators is established.
\end{abstract}

%\keywords{light harvesting complex, incoherent excitation, Lindblad equation, Kraus operator, Lanczos algorithm, Krylov space}

\maketitle

\section{Introduction} \label{sec:introduction}

Light absorbing molecules are ubiquitous in nature, performing the first step of many highly significant biological processes such as vision and photosynthesis via light harvesting complexes (LHCs). Many studies of such molecules have focused on the time dependent evolution of a wavepacket that has been excited by ultrafast laser pulse \cite{dm1,dm2,dm3,dyn1,dyn2,dyn3,dyn4,dyn5,dyn7,mhs,exp1}. However, light absorbing molecules in nature are excited by an incoherent light, which induces a radically different state given by a stationary density matrix~\cite{jiang,brev,dodin19}. This stationary matrix can be obtained perturbatively using the long-time limit of interaction 
with an incoherent light bath \cite{jiang}. However, the perturbative treatment requires use of molecular eigenstates, where the stationary density matrix has a simple diagonal structure. For realistic systems which include many nuclear vibrational degrees of freedom, the diagonalization of the full molecular Hamiltonian is prohibitively expensive to compute. 
Thus, finding efficient computational approaches that circumvent explicit calculation of molecular eigenstates is the main aim of this work. To do so we use the Kraus operator formalism to transform an initial non-stationary density into its stationary counterpart.

This formalism will be extended to consider the LHC as a part of open system with molecular environment as another bath \cite{sst6,dodin19}. Such an extension would give results that are quantitatively dependent on the nature of the system-bath coupling, and is discussed in Appendix~\ref{app:bath}. However, given that the focus of this paper is on developing useful numerical procedures to obtain the coherence-free stationary states, we utilize the artificial bath introduced in Ref.~\citenum{jiang}, i.e. the excited state density matix obtained perturbatively is normalized to avoid the linear-in time growth of population in the excited state due to continuous irradiation by the incoherent light. This normalization corresponds qualitatively to introducing a bath that withdraws energy from the molecular system at the same rate as it is absorbed from the incoherent radiation field. The result is the well-defined stationary state described in Ref.~\citenum{jiang} and utilized below.

This paper is organized as follows. Section~\ref{subsec:pt} follows the derivation of Ref.~\citenum{jiang} for obtaining the stationary density matrix from the long-time limit of incoherent light-matter interaction. Section~\ref{subsec:kraus} offers a short review of the Kraus operator formalism and connects it to generating the stationary density matrix. Section~\ref{subsec:methods} establishes three computationally efficient methods for obtaining the stationary density matrix. Two numerically exactly solvable models that incorporate the main features from LHCs are introduced in Sec.~\ref{sec:results}, and the convergence of all methods is studied. Finally, Sec.~\ref{sec:summary} offers a summary and outlook. Atomic units are used throughout the paper.

\section{Theory} \label{sec:theory}

\subsection{Long-time incoherent-light excitation} \label{subsec:pt}
Here we will provide the derivation for incoherent, long-time excitations~\cite{jiang} that play a pivotal role in this work.

Consider a molecular Hamiltonian $\hat H$ containing nuclear and electronic degrees of freedom 
with discrete energy eigenstates $\ket{E_i}$ and energies $E_i$. This Hamiltonian is coupled to a 
light field through the transition dipole operator $\hat\mu$ and the electric field $\epsilon(t)$
\begin{equation}
\hat H_{LM} = \hat H + \hat V_{LM} = \sum_i E_i \ket{E_i}\bra{E_i} - \hat\mu [\epsilon(t)+\epsilon^*(t)],
\end{equation}
where for simplicity, $\hat\mu$ is considered to be aligned with the electric field. Using the rotating wave approximation, and considering the weak-coupling limit, first-order perturbation theory yields the wavefunction at time $t$
\begin{equation}
\ket{\phi(t)} = - i \sum_j\mu_{j}\int_{-\infty}^t d\tau \epsilon(\tau)e^{i\omega_{j0}\tau}\ket{E_j}e^{-iE_j t},
\end{equation}
where we used the ground state of the molecule $\ket{E_0}$ as the initial state, $\mu_{j}\equiv\bra{E_j}\hat\mu\ket{E_0}$ and $\omega_{jk}\equiv E_j-E_k$. Since we are considering an incoherent light source, the electric field $\epsilon(t)$ cannot be described analytically, but requires an ensemble averaging due to its stochastic nature. This averaging leads to a mixed density matrix $\rho_\phi(t)$ description of the molecular state. $\rho_\phi(t)$ is obtained from the product $\ket{\phi(t)}\bra{\phi(t)}$ by averaging over the light statistics. Since the duration time of the excitation (i.e. solar irradiation) is orders of magnitude larger than all the other significant timescales of the process, we are interested in the limit $t\rightarrow\infty$ for the density matrix
\begin{align} 
\rho_\infty &\equiv \rho_\phi(t\rightarrow\infty) = \langle\ket{\phi(\infty)}\bra{\phi(\infty)}\rangle_\Gamma \nonumber \\
&= \sum_{k,j} \ket{E_k}\bra{E_j} \mu_k\mu_j^*\int_{-\infty}^\infty d\tau_1\int_{-\infty}^\infty d\tau_2  \nonumber \\
&\times e^{-i\omega_{k0}\tau_1} e^{i\omega_{j0}\tau_2} \langle\epsilon(\tau_1)\epsilon^*(\tau_2)\rangle_\Gamma , \label{eq:pre_rho}
\end{align}
where $\langle \odot \rangle_\Gamma$ denotes averaging over the light source statistics. As discussed in Refs.~\citenum{chenu,jiang,brev}, only the first-order correlation function of the light source plays a role in the density matrix. Using the first-order correlation function for thermal light \cite{optics1,optics2}:
\begin{align} \label{eq:G1}
\langle\epsilon(\tau_1)\epsilon^*(\tau_2)\rangle_\Gamma &= G^{(1)}(\tau_1,\tau_2) \nonumber \\
&= \frac{2}{3\pi}\int_0^\infty d\omega \omega^3 n(\omega)e^{-i\omega(\tau_1-\tau_2)},
\end{align}
with $n(\omega)=(e^{-\beta\omega}-1)^{-1}$ the average photon number at temperature $T=(\beta k_B)^{-1}$ and $k_B$ the Boltzmann constant, we define the light intensity spectrum as $I(\omega)\equiv \omega^3n(\omega)$. Using these results in Eq.~(\ref{eq:pre_rho}) yields the normalized density matrix
\begin{align}\notag
\rho_\infty &\propto \sum_{k,j} \ket{E_k}\bra{E_j} \mu_k\mu_j^*\int_{-\infty}^\infty d\tau_1 \int_{-\infty}^\infty d\tau_2 \\
&\times e^{-i\omega_{k0}\tau_1}e^{i\omega_{j0}\tau_2}\int_0^\infty d\omega I(\omega) e^{i\omega(\tau_1-\tau_2)},
\end{align}
where a proportionality constant enforces $\Tr{\rho_\infty}=1$. 
Changing the integration order and evaluating the terms in the exponentials reduces this expression to
\begin{equation}
\rho_\infty \propto \sum_{k,j} \ket{E_k}\bra{E_j} \mu_k\mu_j^* \int_0^\infty d\omega I(\omega) \delta(\omega_{j0}-\omega) \delta(\omega_{k0} - \omega).
\end{equation}
The delta terms eliminate all the off-diagonal terms, obtaining a stationary density matrix which can be written as
\begin{equation} \label{eq:incoherent}
\rho_\infty \propto \sum_j \abs{\mu_j}^2 I(\omega_{j0}) \ket{E_j}\bra{E_j}.
\end{equation}
The density matrix consists of two parts: $I(\omega)$ power spectrum of the source, and 
$\abs{\mu_j}^2$ factors containing Franck-Condon overlap integrals between the ground and excited molecular states. Defining $\hat L = \sum_k \sqrt{I(\omega_{k0})}\ket{E_k}\bra{E_k}$, we obtain a modified Franck-Condon wavefunction $\ket{\tilde \mu} \propto \hat L \hat\mu \ket{E_0}$.
Details for building $\hat L$ from $\hat H$ in a computationally stable and efficient way can be found in Appendix~\ref{app:L}. For cases where the linear absorption spectrum of the molecule is localized over a small energy interval or where $I(\omega_{j0})$ is essentially constant, $\hat L$ will have a very similar action on all populated eigenstates, meaning $\ket{\tilde\mu}$ will be very similar to a ``Franck-Condon wavefunction'' $\hat \mu\ket{E_0}$. 

Finally, the incoherent density for long-time incoherent excitations from Eq.~(\ref{eq:incoherent}) can be written as
\begin{equation} \label{eq:rho}
\rho_\infty = \sum_j \ket{E_j}\bra{E_j}\rho_{\tilde\mu}\ket{E_j}\bra{E_j}.
\end{equation}
where the pure density matrix 
\begin{align}
\rho_{\tilde\mu} &\equiv \ket{\tilde\mu}\bra{\tilde\mu} \nonumber \\
& \propto \sum_{k,j} \sqrt{I(\omega_{k0})I(\omega_{j0})} \mu_k\mu_j^*\ket{E_k}\bra{E_j}
\end{align}
corresponds to an ultrafast Franck-Condon excitation which has been modified to include the light spectrum.  $\rho_{\tilde\mu}$ will serve as an initial approximation for all proposed approaches to obtaining the 
stationary density matrix $\rho_\infty$. Following the perturbation theory route to $\rho_{\tilde\mu}$ requires long-time propagation and thus we consider a more efficient approach based on Kraus operators instead. 

\subsection{Kraus operators and decoherence} \label{subsec:kraus}
In general, a quantum operation $\Phi$ that maps physical density matrices (i.e. those that are normalized, non-negative, and Hermitian),
also known as \gls{CPTP} map, can always be expressed as
\begin{equation} \label{eq:kraus_channel}
\Phi[\rho]= \sum_k \hat J_k \rho \hat J_k^\dagger,
\end{equation}
where $\hat J_k$ are Kraus operators. Equation (\ref{eq:rho}) is already in the Kraus form, where the Kraus operators correspond to 
projectors $\ket{E_j}\bra{E_j}$. It has been shown that the same quantum operation can be represented using different operators $\hat J_k$ \cite{kraus3,kraus4,kraus5}, meaning that the set of operators $\lbrace \hat J_k\rbrace$ is not unique for a given operation. 
This is due to the so-called issue of unraveling the ensemble \cite{ensemble}, which states that there is an infinite number of different ensembles 
(or sets of pure states) giving rise to the same density matrix. This can be seen as a lack of injectivity in the ensemble to 
density matrix relationship. In our case, the quantum operation completely decoheres an initial pure density 
$\rho_{\tilde\mu}$ to the energy basis, and the unraveling issue translates to having many different ways of building a decohering procedure.
(General considerations regarding decoherence to the energy eigenstate basis are discussed in Ref.~\citenum{zurek}). Our focus will be on computationally efficient ways to achieve this decoherence.  

We will consider two types of approaches when building the \gls{CPTP} mapping. The first one explicitly builds the Kraus operators by approximating the projectors $\hat J_k = \ket{E_k}\bra{E_k}$, which would correspond to obtaining the density matrix as a set of pure eigenstates. In the second type, the general form for our approximations will involve $t$-parameter dependent procedures 
for decohering a Franck-Condon excitation
\begin{equation} \label{eq:functional}
\rho(t) = \sum_{k,j} f(E_j,E_k,t) \ket{E_k}\bra{E_k} \hat\mu \ket{E_0}\bra{E_0}\hat\mu \ket{E_j}\bra{E_j},
\end{equation}
where $\lim_{t\rightarrow\infty} f(E_j,E_k,t) = \delta(E_k - E_j)I(\omega_{k0})$, but $t$ is not necessarily physical time. 
If the initial density matrix is $\rho_{\tilde\mu}$, which already includes the information of the spectrum $I$, 
then the long-time limit requirement relaxes to only the delta part of the expression.

\subsection{Stationary state generation} \label{subsec:methods}

\subsubsection{Dynamic averaging} \label{subsubsec:time}

Averaging density matrices at different times allows one to obtain the stationary density in the long-time limit
\begin{equation} \label{eq:dyn_kraus}
\rho_t = \frac{1}{t}\int_{0}^{t} d\tau \hat U_\tau \rho_{\tilde\mu} \hat U_\tau^\dagger,
\end{equation}
where $\hat U_\tau = \exp[-i\hat H \tau]$.
This expression is already in a Kraus form, having a continuous sum over $\tau$ with Kraus operators $\hat J_\tau^{(t)} = \hat U_\tau/\sqrt{t}$. Using the eigenbasis representation of the operators and performing the time integration yields an enlightening expression:
\begin{equation} \label{eq:dyn_explicit}
\rho_t = \sum_{k,j}\textrm{sinc}\Big(\frac{\omega_{kj}t}{2}\Big)e^{\frac{-i\omega_{kj}t}{2}} \ket{E_k}\bra{E_k}\rho_{\tilde\mu}\ket{E_j}\bra{E_j},
\end{equation}
where $\textrm{sinc}(x) = \sin(x)/x$. When $t\rightarrow\infty$, $\textrm{sinc}(\frac{\omega_{kj}t}{2})\rightarrow\delta(\omega_{kj})$, showing that $\rho_t\rightarrow\rho_\infty$. 

This dynamical approach corresponds to a completely quantum propagation, as opposed to, e.g., the recently proposed mixed quantum-classical method for solar excitations \cite{barbatti}. As such, this dynamical method only requires one propagation, as opposed to the ensemble of trajectories necessary for a quantum-classical approach. \\

This approach achieves decoherence by considering the stationary density matrix as a mixture of coherent wavepackets, all having the same eigenstate populations but different phases (see also \cite{chenu}). We will now consider a more traditional approach to decoherence through the use of a master equation.

\subsubsection{Lindbladian decoherence} \label{subsubsec:lindblad}
The Lindblad equation is the simplest quantum master equation that can describe decoherence and energy dissipation caused by a Markovian environment while remaining completely positive-definite \cite{lindblad,qinfo}. The Lindblad equation is generally written as
\begin{equation} \label{eq:gen_lind}
\dot \rho = -i[\hat H,\rho] + \gamma \sum_{j} \big( \hat L_j\rho \hat L_j^\dagger -\frac{1}{2} \lbrace \hat L_j^\dagger\hat L_j,\rho \rbrace \big),
\end{equation}
where the first term on the right hand side represents the unitary evolution of the system, and the second term is the Lindbladian that can induce decoherence and energy dissipation due to the interaction with the environment. We will use a Lindblad-like equation 
\begin{align} \label{eq:lindblad}
\frac{d\rho_L(\tau)}{d\tau} &= - [\hat H,[\hat H,\rho_L(\tau)]],
\end{align}
as a way to decohere an initially pure state $\rho_L(\tau=0) = \rho_{\tilde\mu}$, while conserving all populations of molecular eigenstates. This nested commutator recovers the Lindbladian structure if we chose, in Eq.(\ref{eq:gen_lind}), $\hat L_j = \hat H$ and $\gamma = 2$. Here, $\tau$ is a variable devoid of physical meaning that serves to evolve the equation. Choosing a different $\gamma$ will change the convergence rate with respect to $\tau$ while also modifying the stiffness of the equation, making the computational effort independent of $\gamma$.

Equation~\eqref{eq:lindblad} becomes more transparent if we consider the superoperator picture, 
where this nested commutator corresponds to an infinitesimal generator for the group of purely decohering operations. 
Solving in the energy eigenbasis representation yields the result
\begin{equation} \label{eq:lindblad_gaus}
\rho_L(\tau) = \sum_{k,j} e^{-\omega_{kj}^2 \tau} \ket{E_k}\bra{E_k}\rho_{\tilde\mu}\ket{E_j}\bra{E_j}.
\end{equation}
This solution has a gaussian convergence to the incoherent density with a rate given by the energy difference between states squared. 
Equation \eqref{eq:lindblad} can also be solved using the Kraus representation \cite{kraus2}, obtaining a Kraus operator form:
\begin{equation}
\rho_L(\tau) = e^{-\hat H^2\tau}\sum_{k=0}^\infty \frac{(2\tau)^k}{k!}\hat H^k \rho_{\tilde\mu} \hat H^k e^{-\hat H^2\tau}.
\end{equation}
Computational evaluation of this sum has a poor convergence with $k$ except for very small $\tau$'s, rendering the explicit version of this approach unfeasible. \\

There are approaches for obtaining the long-time limit state to which a Lindbladian equilibrates, avoiding the need to perform dynamics \cite{lindblad}. However, such approaches rely on the symmetries of the Lindblad operator interaction terms. Since in our case these terms include the Hamiltonian operator, directly obtaining the long-time limit necessitates the full diagonalization of the Hamiltonian, which is unfeasible for realistic molecular systems. Therefore, we evolve the density matrix following Eq.~(\ref{eq:lindblad}).

\subsubsection{Lanczos shift-and-invert steps} \label{subsubsec:lanczos}
From Eq.~(\ref{eq:functional}), the function $f$ depends on the energies. This means that the previously mentioned methods can also be represented using a polynomial expansion of the Hamiltonian, as seen in the explicit version of the Lindblad method discussed in the previous section. Although the computational evaluation of these explicit polynomial expansions is not convergent, it should be clear that the Kraus operators require information about the application of powers of the Hamiltonian to $\ket{\tilde\mu}$: the decoherence is in the energy basis, thus the only information required is how any function of $\hat H$ operates on $\ket{\tilde\mu}$. 

The Krylov subspace is created by the sequential application of $\hat H$ to an initial wavefunction, also called the seed vector, $\ket{\tilde\mu}$: $\mathcal{K}_n(\ket{\tilde{\mu}})=\lbrace\ket{\tilde\mu},\hat H \ket{\tilde\mu},\hat H^2\ket{\tilde\mu},...,\hat H^n\ket{\tilde\mu}\rbrace$. Its construction is not computationally stable unless its vectors are orthonormalized at each step.  Due to 
hermiticity of $\hat H$, its matrix represented in the orthonormal Krylov space is tridiagonal \cite{linear1,linear2}. Commonly known as a Lanczos procedure, diagonalization of the tridiagonal matrix 
at step $n$ yields $n$ approximate eigenvectors $\lbrace\ket{r_k^{(n)}}\rbrace_{k=1}^n$ (Ritz vectors). From this, Kraus operators can 
be built as
\begin{equation} \label{eq:kraus_ritz}
\hat J_k^{(n)} \propto \ket{r_k^{(n)}}\bra{r_k^{(n)}},
\end{equation}
where a proportionality constant can be found from $\tr{\Phi[\rho_{\tilde\mu}]} = 1$, where
$\Phi$ is defined in Eq. (\ref{eq:kraus_channel}). 
Note that the Kraus map based on $\hat J_k^{(n)}$ is not \gls{CPTP} for an arbitrary starting physical density matrix. Yet, 
for the starting $\rho_{\tilde\mu}$ it does describe a \gls{CPTP} operation, which is sufficient for our purpose. 
The basic idea of this approach is to approximate some eigenstates, and use their associated projectors as Kraus operators. 
The practically relevant question becomes how to efficiently approximate those eigenstates that are most important for decoherence 
of $\rho_{\tilde\mu}$.

When building the quantum operation, only the Ritz vectors that approximate eigenvectors for which $\langle E_k | \tilde\mu\rangle \neq 0$ 
will have a non-zero contribution when considering the Kraus operator sum. 
Generally, Ritz vectors quickly converge for extremal parts of the spectrum but not for interior eigenvalues \cite{lanczos_conv}.
Since the energy eigenstates that build a Franck-Condon excitation are usually deep inside the energy spectrum, 
a shift-and-invert variant of the Lanczos procedure yields faster convergence. By using a spectral transformation of the Hamiltonian, 
its eigenvectors are conserved while modifying the spectrum \cite{lanczos1}. That is, the Lanczos procedure applied with 
$(\hat H-\sigma\hat 1)^{-1}$ instead of $\hat H$ yields a quick convergence of the Ritz 
vectors to the eigenstates closest to $\sigma$. We chose $\sigma = \langle \tilde\mu \vert\hat H \vert \tilde\mu\rangle$, 
which becomes an increasingly accurate criterion as the size of the Hamiltonian matrix grows. 
Further implementation details of the shift-and-invert procedure are provided in Appendix~\ref{app:lanczos}.

\section{Model systems and computational results} \label{sec:results}

We consider two low-dimensional model systems where one can simulate a nontrivial interplay between electron and nuclear degrees of freedom. 
The small number of degrees of freedom for these models allows us to solve them exactly and monitor rates of convergence for the methods described above.

\subsection*{1D-LVC model}

\begin{figure}
	\centering
	\includegraphics[width=8cm]{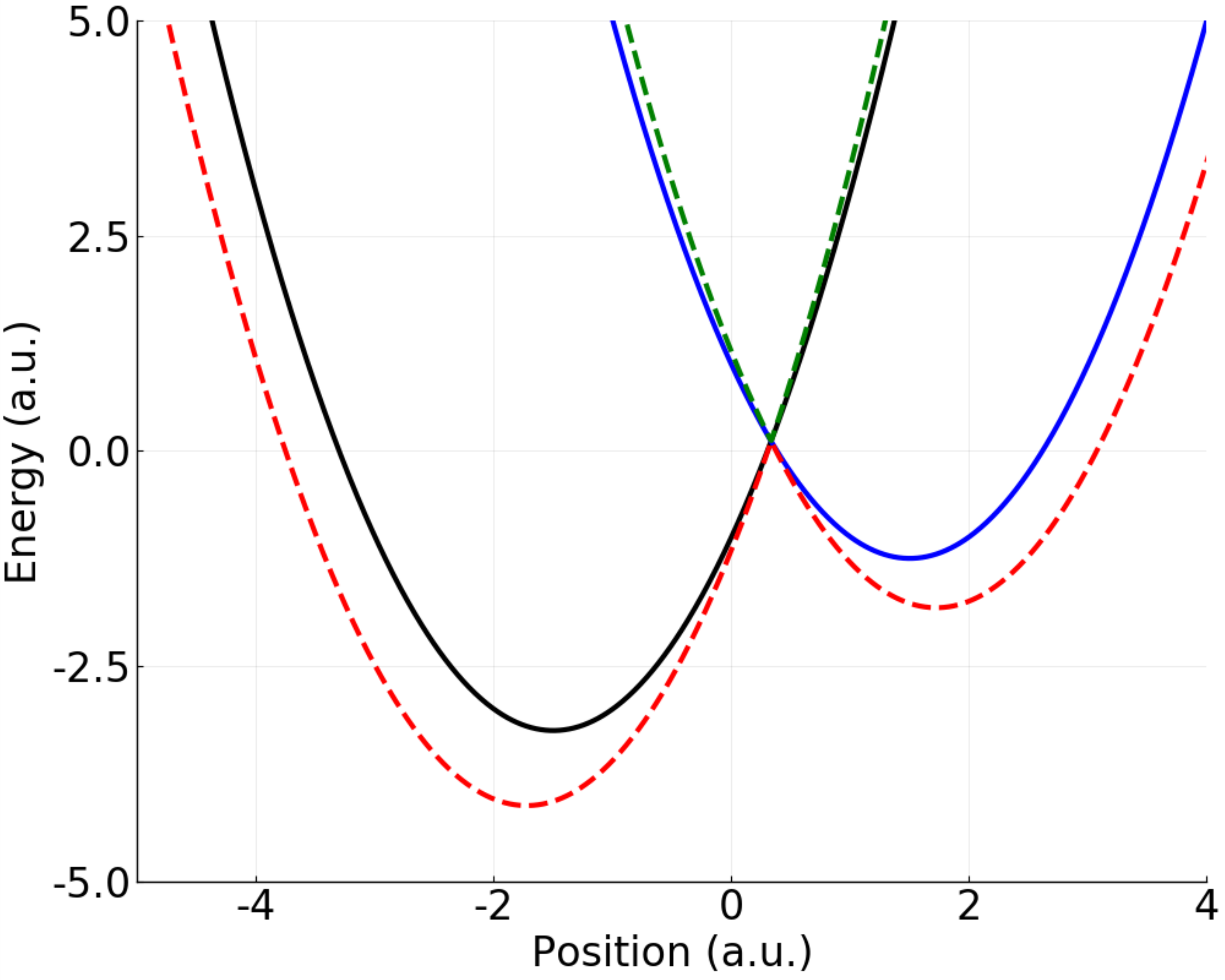}
	\caption{Diabatic (solid) and adiabatic (dotted) potential energy surfaces for the 1D-LVC model.}
	\label{fig:1D-PES}
\end{figure}

The one-dimensional \gls{LVC} model (Fig.~\ref{fig:1D-PES}) having donor and acceptor harmonic wells, which are coupled through a linear vibronic term, is written as

\begin{align}
\hat H_{1D} &= \frac{\omega}{2}\Big(q^2 - \frac{\partial^2}{\partial q^2}\Big)\hat 1 \nonumber \\
&\ + \left[\begin{array}{cc}
aq-\Delta/2 & c(q-\frac{\Delta}{2a}) \\
c(q-\frac{\Delta}{2a}) & \Delta/2 - aq
\end{array}\right],
\end{align}
where $\omega=2$ is the frequency of the harmonic oscillator, $q$ is the corresponding dimensionless normal-mode coordinate, $\Delta=2$ is the energy difference between the ground and excited electronic states, $c=1.7$ is the diabatic coupling and $a=3$ corresponds to the displacement in the position of the acceptor well. 
We used $30$ harmonic oscillator basis functions for the two states, which results in $60\times60$ matrix of $\hat H_{1D}$. 

\subsection*{2D-retinal model}

\begin{figure}
	\centering
	\includegraphics[width=8cm]{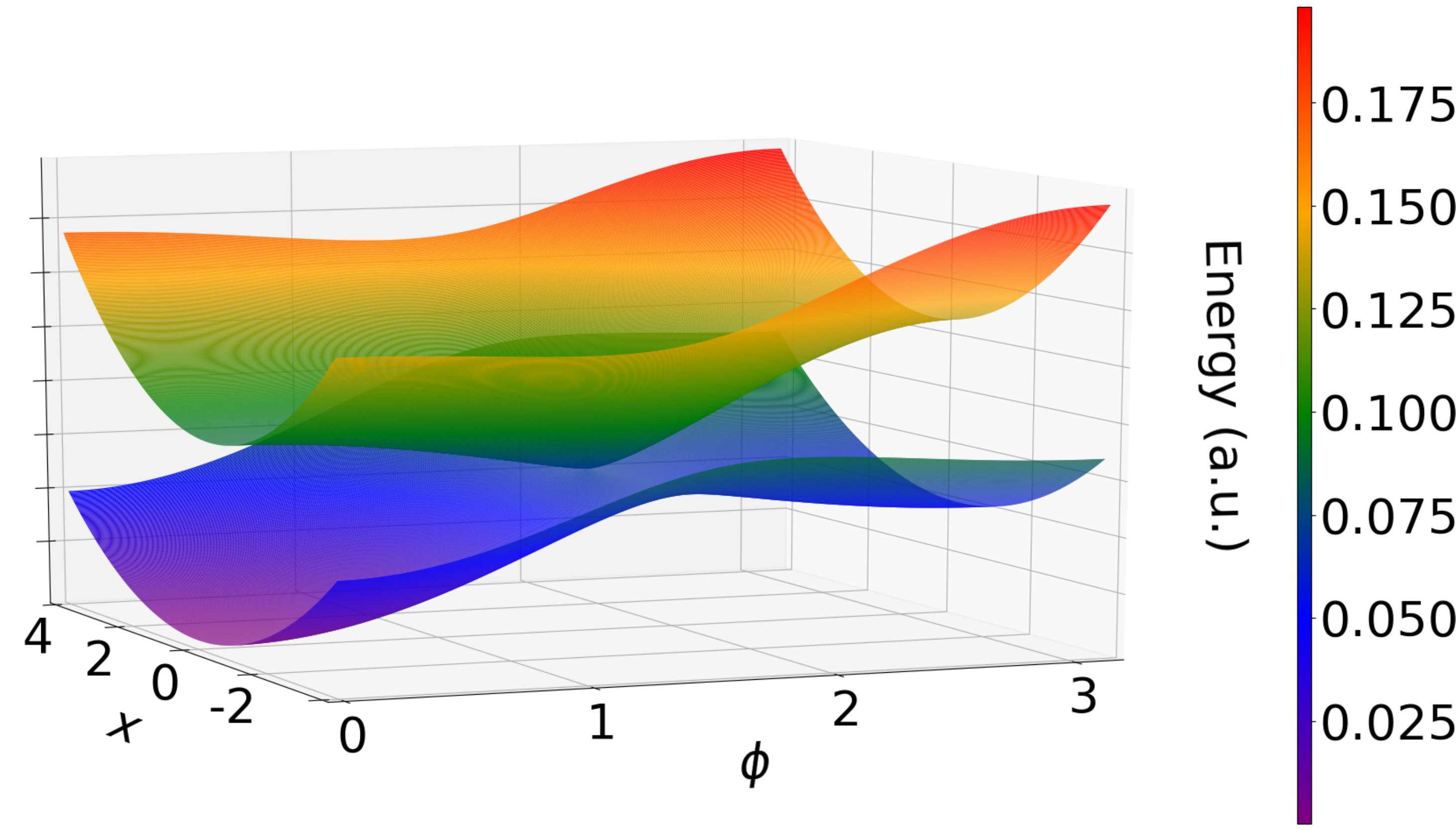}
	\caption{Adiabatic potential energy surfaces for the retinal model.}
	\label{fig:ret-PES}
\end{figure}

The two-dimensional Hahn and Stock retinal model \cite{hs} was originally introduced to reproduce salient features of the retinal molecule while remaining computationally tractable. The two nuclear degrees of freedom correspond to the torsional coordinate $\phi$, 
whose motion gives the {\it cis} to {\it trans} isomerization, and a coupling coordinate $x$, which corresponds to high-frequency 
non-reactive modes. The vibronic Hamiltonian is defined as
\begin{eqnarray}
\hat H_{HS} &=& \hat T \hat 1 + \begin{bmatrix}
E_0 + \frac{\omega}{2}x^2 & \lambda x \\
\lambda x & E_1 + \frac{\omega}{2}x^2 + \kappa x
\end{bmatrix}
\nonumber \\
&& + \begin{bmatrix}
\tilde V_0(1-\cos\phi) & 0 \\
0 & - \tilde V_1(1-\cos\phi)
\end{bmatrix}, \label{eq:hs}
\end{eqnarray}
where $\hat T = -(1/(2m))(\partial^2/\partial\phi^2) - (\omega/2)(\partial^2/\partial x^2)$ is the kinetic energy operator, and the other terms are 
the interaction potential in a basis spanned by diabatic electronic functions $\ket{\psi_n}$, with $n=0,1$. The model parameters were 
chosen to reproduce the femtosecond dynamics of retinal in rhodopsin \cite{hs}, and are as follows (in a.u.): 
$E_0 = 0$, $E_1 = 9.11\times 10^{-2}$, $\tilde V_0 = 6.61\times 10^{-2}$, $\tilde V_1 = 4.01\times 10^{-2}$, $\omega = 6.98\times 10^{-3}$, $\kappa = 3.67\times 10^{-3}$, $\lambda = 6.98\times 10^{-3}$ and 
$m^{-1} = 1.78\times 10^{-5}$. Figure \ref{fig:ret-PES} shows the adiabatic potential energy surfaces for this model. 
We used $20$ harmonic oscillator basis functions for the coupling coordinate $x$. Since the system Hamiltonian is an even function of $\phi$ and does not couple even and odd functions, only even functions are necessary. We used the first $200$ even eigenfunctions (i.e. having the lowest energies) for the torsional coordinate $\phi$. This was done for both electronic states, resulting in a matrix of $8000\times8000$ for $\hat H_{HS}$.

\subsection*{Quantities of interest}

To benchmark our approaches we evaluate three different quantities: 
1) the purity $\tr{\rho^2}$, which decreases for the stationary density matrix; 2) the $S_0$ electronic state population 
$\tr{\hat P_{S_0} \rho}$, which allows us to assess convergence of observables; and 3) the nuclear \textit{trans} population for the retinal model  $\tr{\hat P_{trans}^{(1)}\rho}$, which is one of the quantities used to define the quantum yield \cite{dodin19}, where
\begin{equation}
\hat P_{trans}^{(1)}=\Theta(\abs{\phi}-\pi/2) \ket{\psi_1}\bra{\psi_1}.
\end{equation}
Here, $\Theta(x)$ is the Heaviside step function, $\phi$ is the nuclear torsional coordinate for the retinal model (see Fig.\ref{fig:ret-PES}), and $\ket{\psi_1}$ is the diabatic electronic function corresponding to the state $S_1$.

Since we are evaluating the ability of the methods to remove coherences from an initially pure density matrix, we worked with $\rho_{\mu}\propto \hat\mu \ket{E_0}\bra{E_0}\hat\mu^\dagger$ as the initial state, which corresponds to Franck-Condon excitation. By building the $\hat L$ operator as described in Appendix~\ref{app:lanczos}, its application after the decoherence procedure also yields the stationary density shown in Eq.(\ref{eq:rho}), making the order in which the $\rho_{\mu}$ is transformed irrelevant.
The evaluation using the dynamical averaging approach requires both the discretization of the integral in Eq.~(\ref{eq:dyn_kraus}), and a dynamical propagation scheme for obtaining $\hat U_\tau \hat\mu\ket{E_0}$. For the latter, the convergence to the incoherent solution does not depend on approximation schemes for the quantum propagator and is determined by Eq.~(\ref{eq:dyn_explicit}). Therefore, $\hat U_\tau$ is built by 
diagonalizing the Hamiltonian. For the integral discretization scheme, we chose a uniform grid with interval $\Delta t$. 
The Lindbladian equation was evolved using an adaptive RK$45$ scheme, which required fewer steps than a fixed step RK$4$ scheme.

For comparing the convergence rates of the different methods we consider the number of steps, which also corresponds to the number of Kraus operators for the dynamic averaging and Lanczos approaches. For the dynamical averaging method, a step 
means adding a new point in the discretization of the integral, while for the Lindbladian approach it is a dynamical step in the Runge-Kutta scheme. 
For the Lanczos approach, each step will give the number of Ritz vectors that are built and considered in the Kraus sum. The benefit of using this criterion is that it is mostly independent of the implementation for each method, allowing us to directly evaluate the efficiency and scalability of the different approaches.

\subsection*{Results}
\begin{figure}
	\centering
	\includegraphics[width=8.5cm]{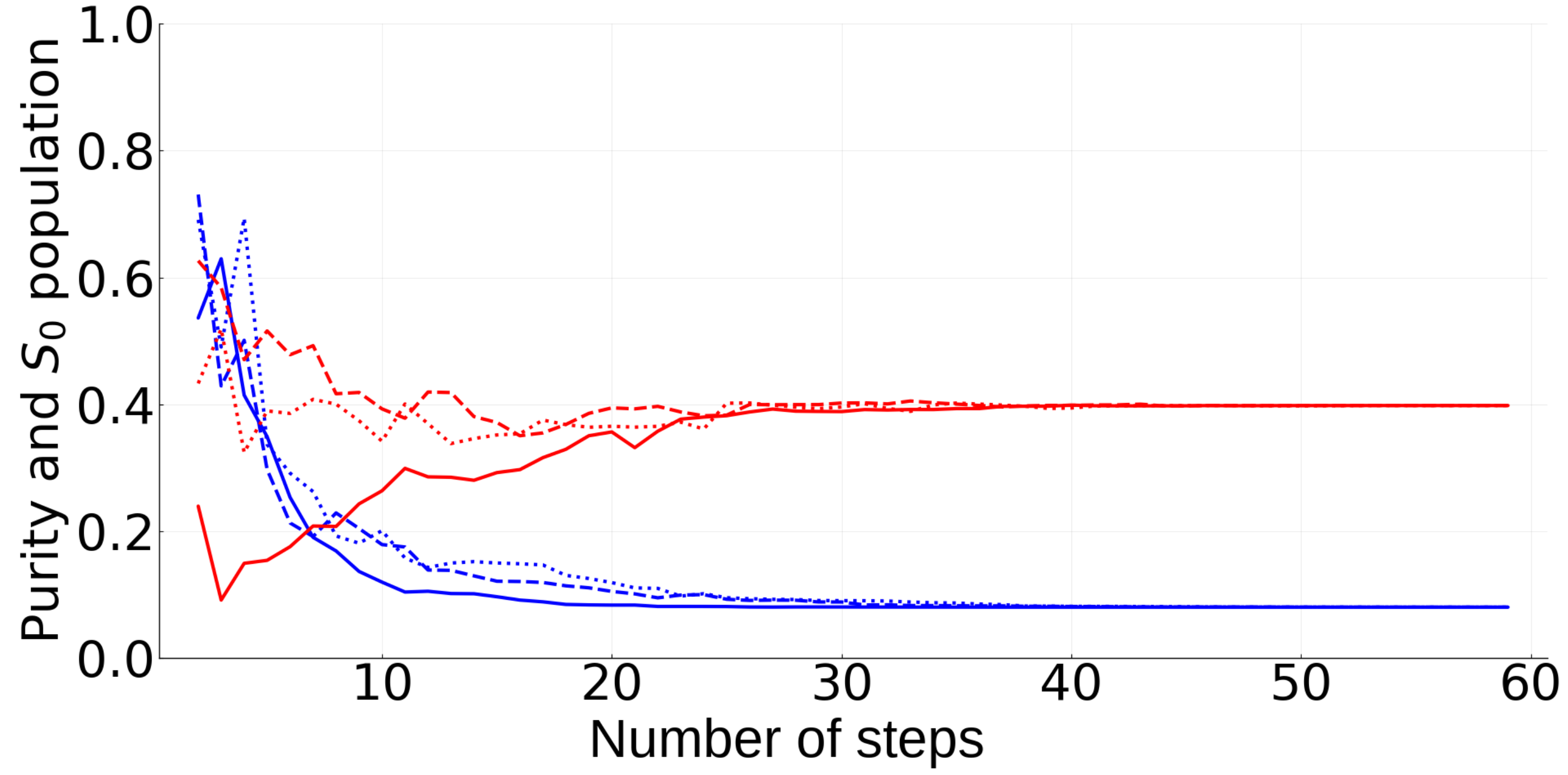}
	\caption{Convergence to purity (blue) and $S_0$ population (red) as a function of the number of Lanczos steps for the \gls{LVC} model. The three different seed vectors discussed in Appendix \ref{app:lanczos} are used, having the Franck-Condon ultrafast excitation $\hat \mu\ket{E_0}$ as the solid line, the corrected excitation $\ket{\mu^{(1)}}$ shown in Eq.\eqref{eq:corr_mu} as the dashed line, and a random seed vector as the dotted line.}
	\label{fig:1D-lanczos}
\end{figure}

\begin{figure}
	\centering
	\includegraphics[width=8.5cm]{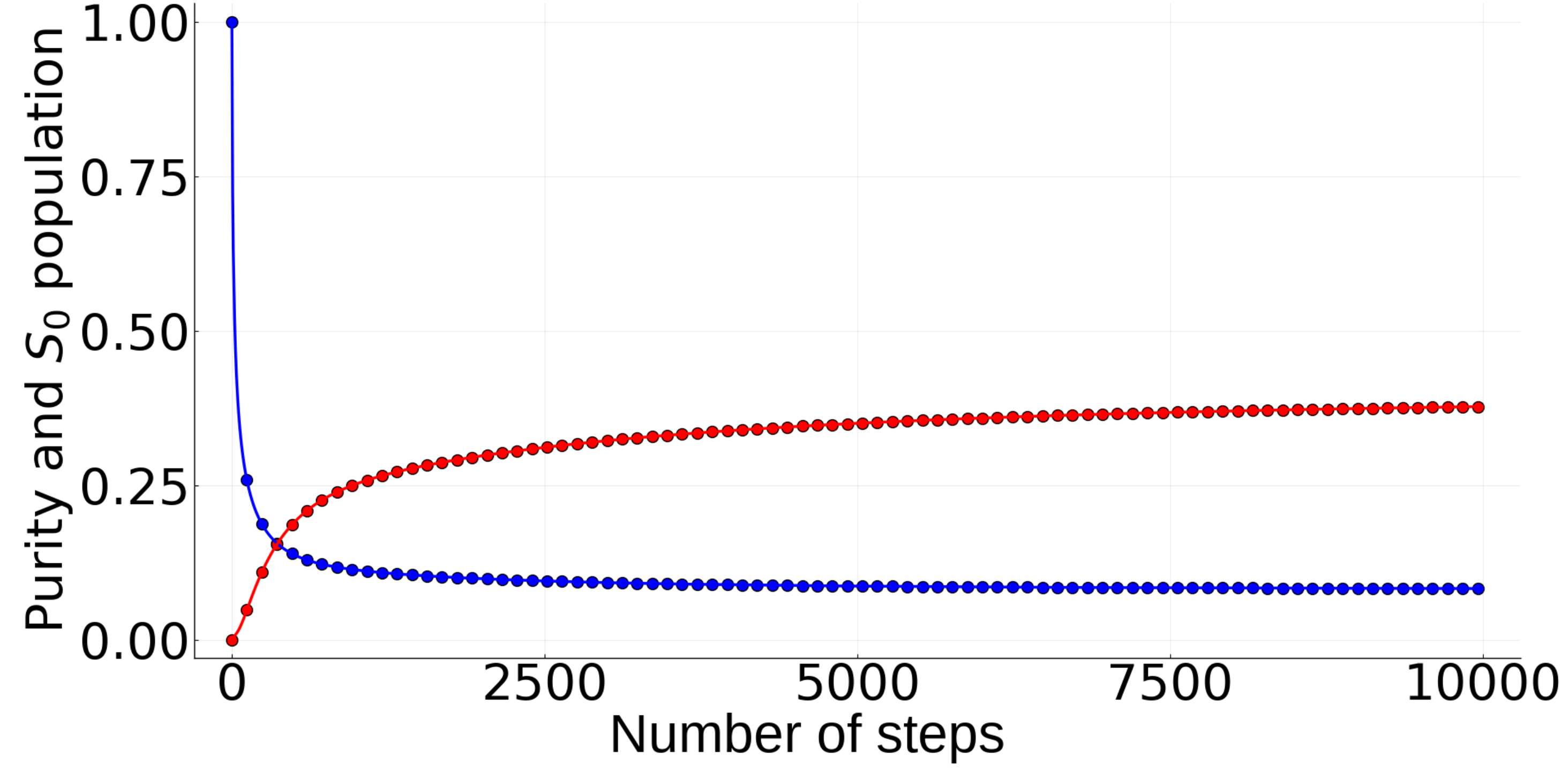}
	\caption{Purity (blue) and $S_0$ population (red) for Lindblad decoherence procedure in the \gls{LVC} model. The dots are the analytical result from Eq.~(\ref{eq:lindblad_gaus}), while the lines are the explicit propagation of the Lindblad equation using the RK$45$ method.}
	\label{fig:1D-lindblad}
\end{figure}

\begin{figure}
	\centering
	\includegraphics[width=8.5cm]{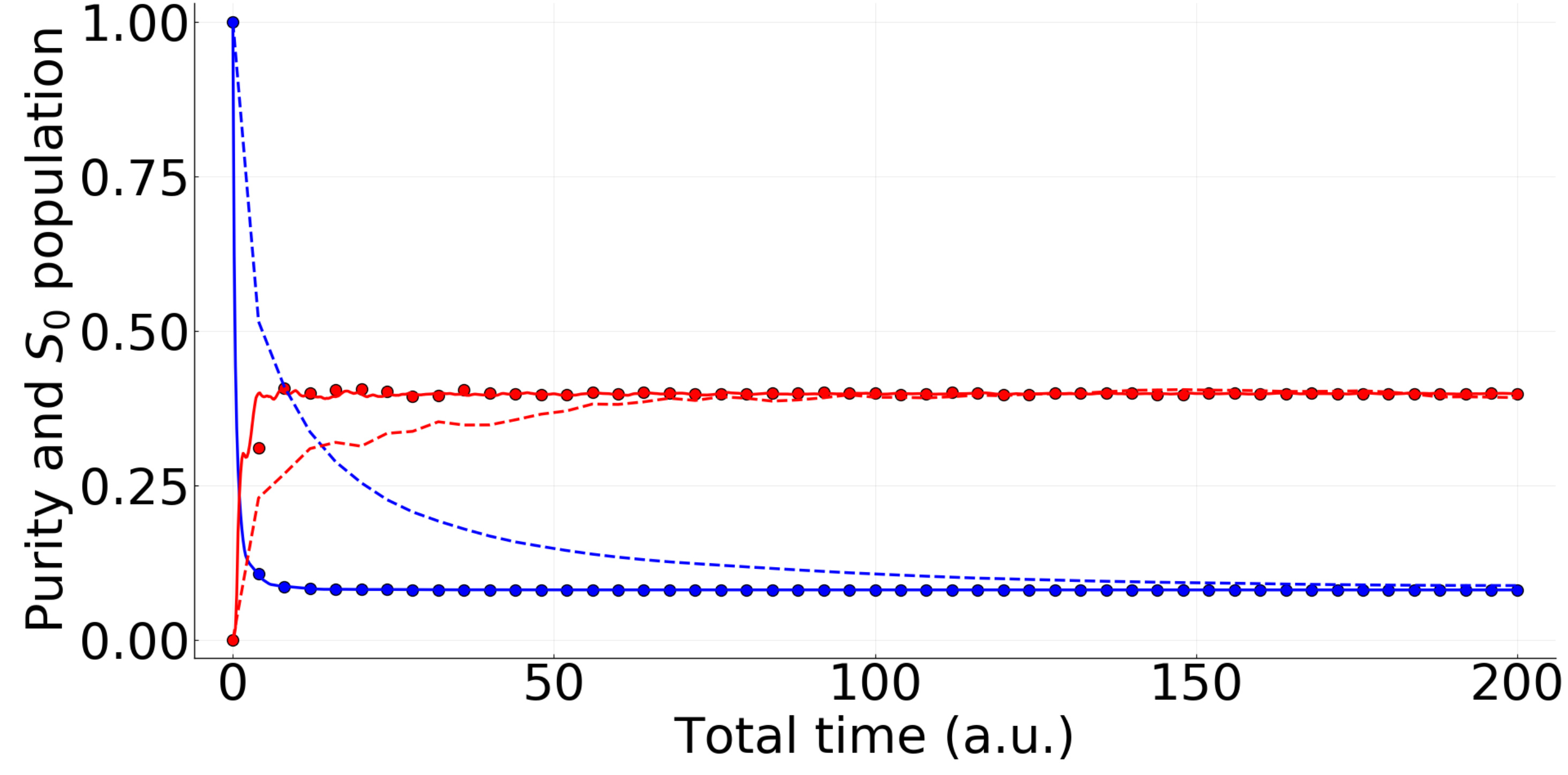}
	\caption{Purity (blue) and $S_0$ population (red) for dynamic averaging approach in the \gls{LVC} model. Dots correspond to the analytical result from Eq.~(\ref{eq:dyn_explicit}). Lines represent the discretized integral, having solid lines for small intervals ($\Delta t = 0.2 \ a.u.$) and dashed lines for large intervals ($\Delta t = 4\ a.u.$).}
	\label{fig:1D-time}
\end{figure}

First, using the \gls{LVC} model, we verify that all methods remove from $\rho_{\mu}$ all coherences in the energy basis with increasing number of steps.
For the Lanczos steps, Fig.~\ref{fig:1D-lanczos} shows the convergence of three different seeds, which all recover the stationary 
density as the number of steps approaches the size of the Hamiltonian basis. The similar behavior of the seed vectors can be explained by the 
small dimensionality of the system, where we quickly recover the complete diagonalization after only $60$ steps. Figure~\ref{fig:1D-lindblad} shows the convergence for the Lindbladian approach, and its propagation exhibits a perfect agreement with the theoretical estimate in Eq.~(\ref{eq:lindblad_gaus}). The dynamic averaging method follows closely its theoretical behavior shown in Eq.~(\ref{eq:dyn_explicit}) when the interval size $\Delta t$ is small, while also recovering the correct limits for the large interval (Fig.~\ref{fig:1D-time}). 
%As expected, all methods approximate the stationary solution as the iterations progress. \\


\begin{figure}
	\centering
	\includegraphics[width=8.5cm]{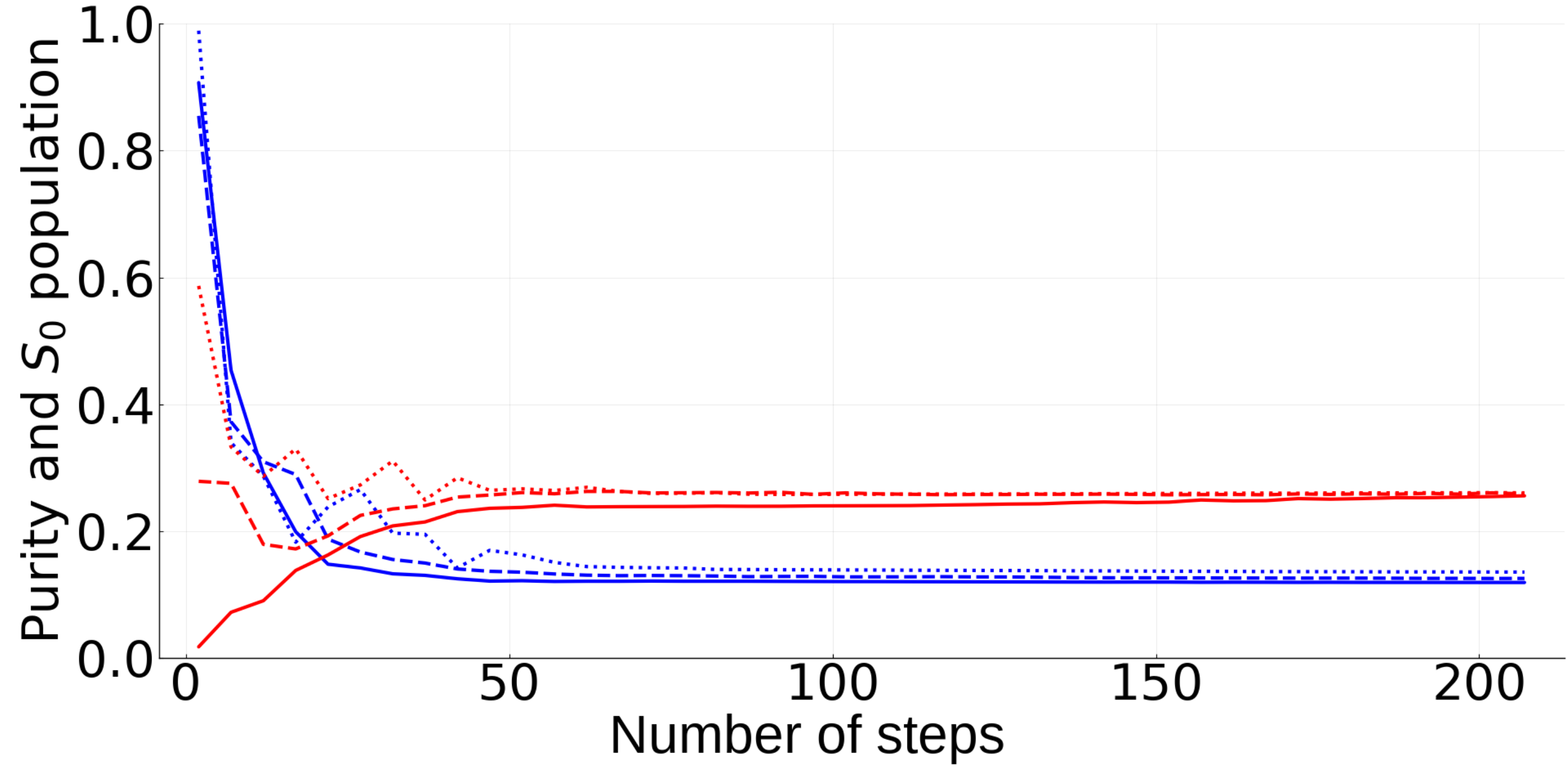}
	\caption{Convergence for the Lanczos method in the retinal model. All legends are identical to those in Fig.\ref{fig:1D-lanczos}.}
	\label{fig:ret-lanczos}
\end{figure}

\begin{figure}
	\centering
	\includegraphics[width=8.5cm]{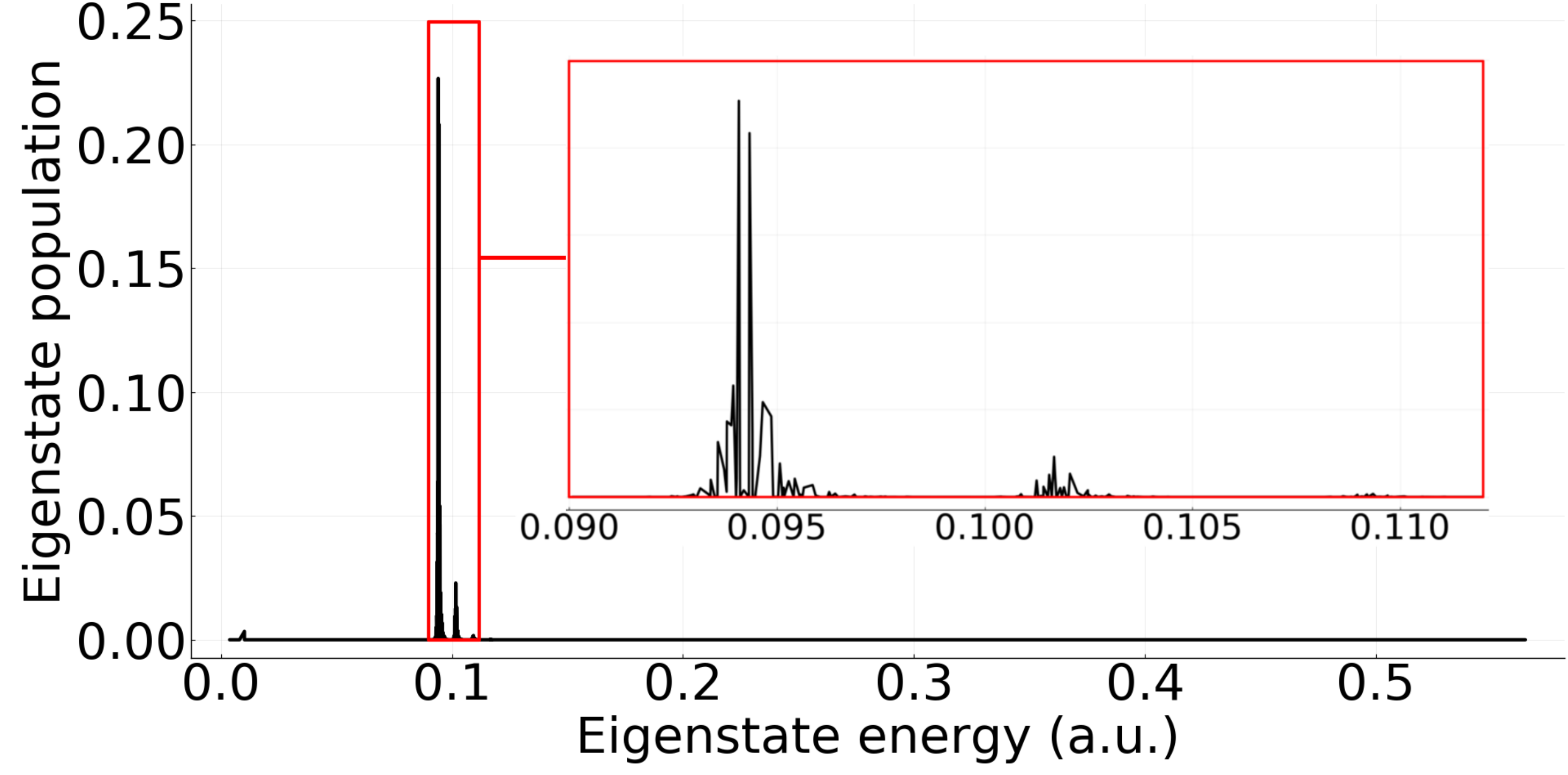}
	\caption{Spectrum of the Franck-Condon excitation in the eigenbasis. 
		There are $369$ eigenstates with energies between $0.09$ and $0.112$ a.u. shown inside the red area.}
	\label{fig:ret-spectrum}
\end{figure}

The Lanczos procedure behavior varies greatly with the seed vector choice when considering larger systems. 
We compare three different seeds for the Lanczos procedure: (1) the ultrafast Franck-Condon 
excitation $\hat\mu \ket{E_0}$, (2) its corrected version $\ket{\mu^{(1)}}$ that is described in Appendix~\ref{app:lanczos}, and (3) a completely random and normalized real vector.
Figure~\ref{fig:ret-lanczos} shows the convergence for all three different seeds in the retinal model, where their different convergence rates become clearly visible. The ultrafast excitation $\hat\mu\ket{E_0}$ has the fastest convergence for purity, followed by the corrected version $\ket{\mu^{(1)}}$. Since both these vectors have mostly components in the energy eigenstates for which $\bra{E_k}\hat\mu\ket{E_0}\neq 0$, the resulting Ritz vectors will quickly approximate the eigenstates of interest, which are close to $\sigma$ and have a non-zero overlap with the ultrafast excitation. This is opposed to the random seed, where the inclusion of most states in the random structure means the generated Ritz vectors will approximate all eigenstates close to $\sigma$, thus generating many states for which $\bra{E_k}\hat\mu\ket{E_0}=0$ and slowing down the convergence for the purity.  Figure~\ref{fig:ret-spectrum} shows the spectrum of $\hat\mu\ket{E_0}$ in the eigenbasis, where we can see many eigenstates with energies close to $\sigma$ and zero overlap with $\hat\mu\ket{E_0}$. When considering the $S_0$ population, the ultrafast excitation clearly has the slowest convergence out of all seeds: it is completely localized in the $S_1$ electronic state, skewing the Lanczos process despite its quick convergence to the eigenstates of interest. Both the random seed and the corrected excitation have a fast convergence to $S_0$ as a result of their balanced populations in both electronic states. The corrected version of the seed achieves a fast convergence for both the $S_0$ population and the purity, which shows it quickly generates the eigenstates of interest while not skewing the Lanczos steps in the $S_0$ populations. \\


\begin{figure}
	\centering
	\includegraphics[width=8.5cm]{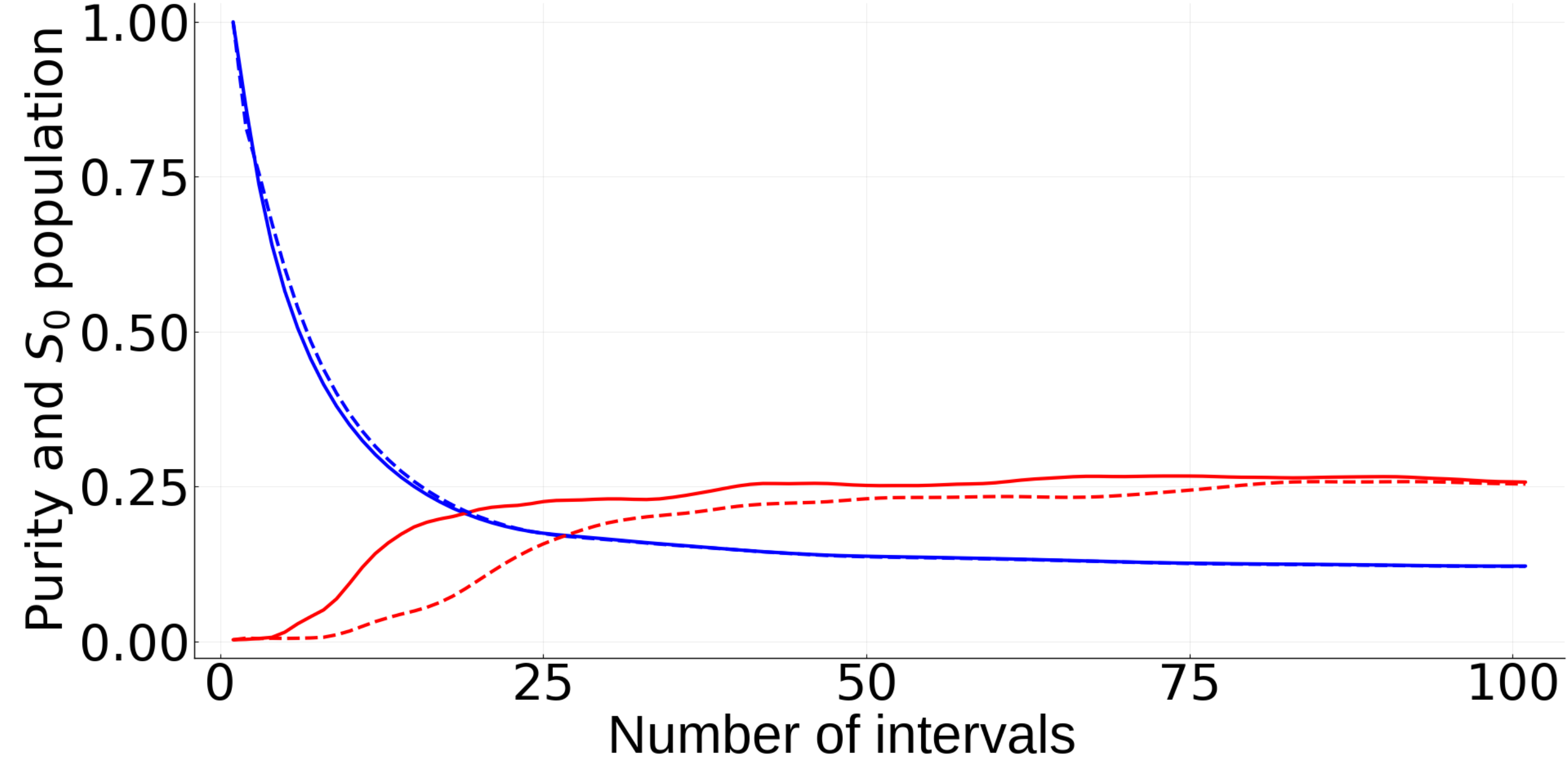}
	\caption{Convergence to purity (blue) and $S_0$ population (red) for dynamic averaging approach in the retinal model. 
		Dashed lines correspond to analytic results from Eq.~(\ref{eq:dyn_explicit}), and full lines to the large interval size $\Delta t = 1000\ (a.u.)$.}
	\label{fig:ret-time}
\end{figure}

The dynamic averaging approach has strong dependence on the discretization of the integral (Fig.~\ref{fig:1D-time}). Only the small interval scheme follows 
the analytic expression of Eq.~(\ref{eq:dyn_explicit}), while the large interval scheme recovers the stationary density only for a large number of steps. 
This can be understood by considering the off-diagonal elements of $\hat U_\tau \rho_{\mu} \hat U_\tau^\dagger$ in the energy basis, 
which oscillate as $e^{\pm i\omega_{kj}\tau}$. The scheme with the small interval size recovers the analytic behavior of Eq.~(\ref{eq:dyn_explicit})
since it accurately approximates the integral. On the other hand, in the scheme with the large interval size, $e^{\pm i\omega_{kj}\tau}$ phases are randomized with respect to each other at each step. The sum of random phases yields an average of zero, recovering the stationary density as the number of steps grows. For the retinal model, the large interval scheme has faster long-time limit convergence than the analytic expression (Fig.~\ref{fig:ret-time}). Since the initial wavefunction $\hat\mu\ket{E_0}$ can be considered a localized statistical fluctuation of a generally delocalized wavefunction $\hat U_\tau\hat\mu\ket{E_0}$,  the small interval size requires long time to ``forget'' the initial state, whereas a large interval places a minimal weight on the initial state, quickly averaging it out after a few steps. This shows that a large $\Delta t$ is beneficial when the size of the Hamiltonian matrix grows. We consider a time large if it allows the wavefunction to change considerably, which can be quantified as $|\bra{E_0}\hat\mu^\dagger\hat U_{\Delta t}\hat\mu\ket{E_0}|^2$ being significantly smaller than $1$ (e.g. $\approx0.3$ for the large interval size in Fig.~\ref{fig:ret-time}). \\ 

For modelling realistic molecular systems, any approach must have a much lower computational cost than a full diagonalization. The retinal model has a matrix dimensionality over $100$ times larger than the \gls{LVC} model, which allows us to compare the efficiency of our approaches for larger matrices. The convergence of all methods is shown in Table~\ref{tab:conv}, where the required number of steps for obtaining purity, $S_0$ and $P_{trans}^{(1)}$ populations within $5\%$ of the exact values are shown. The Lindbladian approach requires more steps than the size of the Hamiltonian matrix, making it prohibitively expensive for realistic systems. The dynamic averaging approach is able to approximate all properties after $84$ iterations, requiring the smallest number of Kraus operators out of all methods. Still, the Lanczos steps also achieved a fast convergence, showing a trade-off for the convergence of purity and the other observables between different seeds. The difference in computational effort between both methods will be mainly given by the inner iterations cost. For the dynamic averaging this translates to a propagation with $\Delta t$ steps, while for the Lanczos procedure it is the linear system solving at each step. For both the Lanczos and dynamical averaging procedures, the required number of steps for a converged density matrix are between $1\%$ and $5\%$ of the Hamiltonian matrix size, showing that both approaches will vastly outperform a full diagonalization if the inner iterations are performed efficiently.

\begin{table}	\caption{Number of steps necessary for reaching a convergence within $5\%$ of the exact value for 
		the different methods in the retinal model. Convergence is considered once the value goes permanently inside $5\%$ range. 
		The Lindbladian approach did not reach convergence after $8000$ steps. For the Lanczos approach with a random seed, $1000$ different seeds were used and their average convergence is considered. The number in parenthesis is the necessary number of additional steps so that at least $99\%$ of the seeds are within the converged range.}
	\centering
	\begin{tabular}{|c|c|c|c|}
		\hline
		Method & Purity & $S_0$ population & $P_{trans}^{(1)}$ \\ \hline
		Dynamic averaging & 84 & 21 & 31 \\
		Lindbladian & >8000 & >8000  & >8000\\
		Lanczos with $\hat \mu\ket{E_0}$ & 50 & 193 & 138 \\
		Lanczos with $\ket{\mu^{(1)}}$ & 267 & 37 & 27 \\
		Lanczos with random & 352 (17) & 45 (14)  & 40 (10)\\ \hline
	\end{tabular}
	
	\label{tab:conv}
\end{table}

\section{Summary and outlook} \label{sec:summary}

In this work we developed methods for generation of the system stationary density matrix originating under incoherent light excitations. 
Such a density matrix can be used for calculating any observable of the system under natural solar light, and it acquires a completely diagonal structure in the energy eigenbasis. Using Kraus operators for modelling a general quantum operation, we showed how this density matrix can be obtained through three different approaches not involving explicit diagonalization.

The first approach performs a dynamical evolution of an initial wavefunction while averaging its associated pure density matrix at different times, recovering a mixed density matrix that approximates the stationary one. We noted how the properties obtained from coherent and incoherent excitations will coincide at short times whenever the linear absorption spectrum is localized over a small energy interval or when the spectrum of the incident light is structureless.

The second approach uses a Lindblad-like equation to induce decoherence in an initially pure state, while the third approach employs a Lanczos shift-and-invert iterative algorithm, building increasingly better guesses to the eigenstates which play a role in the decoherence of the system. The convergence of different seed vectors was studied, showing how an imbalance in their properties might skew the Lanczos process.

The Lindbladian approach suffers from a poor computational efficiency, making it unusable for realistic molecular systems. Both the dynamical averaging and the Lanczos steps behaved more favorably, obtaining accurate approximations of observables with only a fraction of the cost of a full diagonalization. Both these methods offer a computationally efficient and scalable approach for obtaining the stationary density matrix of the system under incoherent light excitations, and the implementation for higher dimensional systems is promising. Scaling of each method with respect to system size, as well as the case of a continuous spectrum, is discussed in Appendix~\ref{app:scaling}.

\section*{Acknowledgements} \label{sec:acknowledgements}
I.L. is grateful to Loïc Joubert-Doriol, Cyrille Lavigne and Ilya Ryabinkin for helpful conversations, and acknowledges the funding of the Anoush Khoshkish Graduate Research Scholarship in Chemistry. A.F.I. and P.B. acknowledge financial support from the US Army Research Office, under grant W911NF-19-1-0433.

\section*{Data availability}
The  code  used  for  the  simulations  and  the  data  that  support the  findings  of  this  study  are  openly  available  in  the  repository \url{https://github.com/iloaiza/Incoherent_density}.

\appendix
\section{Inclusion of second bath} \label{app:bath}
The study adopted here focuses on the interaction of a molecule
with a single incoherent bath, which produces the stationary density
in Eq.~\eqref{eq:incoherent}. Being devoid of off-diagonal coherences, this can be
categorized as an equilibrium state. By contrast,
natural processes, such as photosynthesis, energy transfer
and the first steps in vision, often involve a second
thermal bath, e.g., a protein environment. This results in a 
non-equiibrium steady state (NESS) with non-zero stationary coherences.\cite{dodin19,chern2020}
It is a property of such NESS that they are, most often, independent of
the initial state that generates them. As a consequence, a numerically
useful strategy based upon the work in this paper presents itself.
That is, one can save considerable computational effort by utilizing
the tools developed in this paper to first generate the initial state arising
from the incoherent radiative excitation, and then introducing the second
bath, leading over time to a two-bath NESS. In addition to providing
a useful computational route, this approach allows insight into the
transition from an equilibrium state to a non-equilibrium steady state.
Studies of this type are in progress in our laboratory.

\section{Modified spectrum operator} \label{app:L}
The operator $\hat L$ was defined to include the light spectrum $I(\omega)$ in the initial ultrafast excitation, acting as 
\begin{equation} \label{eq:hat_L}
\hat L \ket {E_k} = \sqrt{I(\omega_{k0})} \ket{E_k}.
\end{equation}
Here, a construction of $\hat L$ using a Chebyshev interpolation scheme is described. \\

Chebyshev interpolation requires the domain of the interpolated function to be in $[-1,1]$. When performing a polynomial expansion using powers of some operator, we can consider the domain to be determined by the spectrum of said operator. The bounds for the spectrum of the Hamiltonian are given by $E_0$ and $E_{max}$, the latter being determined by the basis used for representing $\hat H$. Both of these bounds can be obtained by applying a few Lanczos steps with $\hat H$ over some initial random vector \cite{linear1,linear2}. Once they are obtained, the Hamiltonian is shifted and rescaled by an affine transformation, modifying the bounds of its spectrum from $E_0$ and $E_{max}$ to $-1$ and $1$ respectively:
\begin{equation} \label{eq:affine}
\tilde H = 2\frac{\hat H - E_0\hat 1}{\omega_{max}} - \hat 1,
\end{equation}
having defined $\omega_{max}\equiv E_{max} - E_0$. We now want to find a set of coefficients $\lbrace c_n\rbrace$ so that $\hat L$ can be approximated as a polynomial of $\tilde H$:
\begin{equation} \label{eq:cheby_L}
\hat L \approx \sum_n c_n T_n(\tilde H),
\end{equation}
where $T_n(x)$ is the $n$-th degree Chebyshev polynomial of the first kind. Defining $\tilde\omega_k = \bra{E_k}\tilde H \ket{E_k}$, this quantity is related to $\omega_{k0}$ by the affine transformation (Eq.(\ref{eq:affine})): $\omega_{k0} = \omega_{max}(\tilde\omega_k+1)/2$. This, along with Eq.(\ref{eq:hat_L}), defines the equation for the $c_n$ coefficients:
\begin{equation}
\sum_n c_n T_n(\tilde\omega_k) \approx \sqrt{I\Big(\frac{\omega_{max}(\tilde\omega_k + 1)}{2}\Big)} \equiv L(\tilde\omega_k).
\end{equation}
Obtaining the coefficients $c_n$ is thus equivalent to expanding the function $L(x)$ for $x\in[-1,1]$, which can be done by using a Chebyshev-Gauss quadrature and evaluating it at the Chebyshev nodes \cite{cheby2}:
\begin{equation}
c_n = \frac{2}{N+1}\sum_{j=0}^N L(x_j)T_k(x_j),
\end{equation}
where $N$ is the degree used for the polynomial expansion and $x_j = \cos\Big({\pi(j+\frac{1}{2})/(N+1)\Big)}$ are the roots of $T_{N+1}(x)$. \\

When applying this operator computationally, Eq.~(\ref{eq:cheby_L}) should be calculated using Horner's rule for Chebyshev polynomials (also known as Clenshaw's rule) for computational stability and efficiency \cite{cheby1,cheby2}. This makes the number of applications of $\tilde H$ the same as the degree of the Chebyshev expansion for building $\hat L$. 

\section{Lanczos procedure details} \label{app:lanczos}
\subsection{Invert operation}
The main bottleneck of using a Lanczos shift-and-invert procedure over a regular Lanczos procedure is in the inversion step. 
Inverting the shifted Hamiltonian operator becomes prohibitively expensive as the size of the matrix grows. 
To alleviate this problem, shift-and-invert Lanczos procedures do the inversion by solving a linear system of equations \cite{lanczos1}.
If we consider the shifted operator $\hat H_\sigma = \hat H - \sigma\hat 1$, applying the 
shift-and-invert operator $\hat H_\sigma^{-1}$ on a vector $\ket b$ is equivalent to finding a vector $\ket a$ that solves the linear system
\begin{equation} \label{eq:linear}
\hat H_\sigma \ket{a} = \ket b.
\end{equation}
This linear system is solved at each step of the Lanczos procedure, yielding what is known as an inner-outer algorithm: 
the outer iterations correspond to a full Lanczos step (i.e. application of the operator and orthogonalization), 
while the inner iterations are done by the linear solving procedure for Eq.~(\ref{eq:linear}) at each outer step. 
There is extensive literature on techniques for greatly diminishing the cost of the internal iterations, which can be 
typically achieved using preconditioners and variable tolerances for improving the computational cost without hindering the performance \cite{lanczos1,lanczos2,preco1,preco2,preco3,preco4}. The Lanczos procedure introduced in this paper could greatly benefit from such techniques, 
and its efficient implementation would be crucial for applying the method to realistic molecular systems. 
However, in this study we only seek to showcase the convergence properties of the approach, and the size of the 
model systems allowed us to work explicitly with the shifted and inverted operator, bypassing the need for inner iterations.

\subsection{Corrected seed vector}
Lanczos procedures, as any Krylov subspace method, have a very strong dependence on the seed vector. Generally, the larger the components of the states of interest in the initial seed vector the faster is convergence of the Lanczos procedure. This can be rationalized by considering that the Krylov subspace generated by an initial vector will only contain information on the eigenvectors which have a non-zero overlap with this seed vector, and can be formally justified studying the convergence properties of the Ritz vectors to the eigenvectors \cite{lanczos_conv}. 
This makes random seeds a popular choice for Lanczos procedures: their random character makes them have random components in most if not all states of interest, making sure that in general convergence towards an arbitrary eigenvector is not too slow. 
In principle, this would make the ultrafast excitation $\hat\mu\ket{E_0}$ an ideal choice for a seed vector: the eigenvectors with which it overlaps are the states of interest for the Lanczos procedure, and this results as a rapid convergence for the purity. 

Despite this rapid convergence, this seed suffers from a particularly slow convergence rate for the electronic $S_0$ populations, being outperformed by an initial random seed. If we are interested in a particular observable (e.g. the $S_0$ electronic state population $\hat P_{S_0}$), the seed vector should have 
significant components of the eigenstates of interest for this operator as well. Since the ultrafast excitation has a zero population in the $S_0$ state, the convergence rate for this property is particularly slow. To improve this we propose a corrected seed vector that aims to have components mostly in the eigenstates with $\bra{E_k}\hat\mu\ket{E_0}\neq 0$ while also adding components in the $S_0$ electronic state. Let us consider a diabatic model of a Hamiltonian with two electronic states:
\begin{equation}
\hat H = \begin{bmatrix}
\hat H_{S_0} & \hat V^\dagger \\
\hat V & \hat H_{S_1}
\end{bmatrix} \equiv \hat{\mathcal{H}}_0 + \hat{\mathcal{V}},
\end{equation}
where $\hat H_{S_i}$ are the nuclear Hamiltonians for electronic states $S_i$ respectively, and $\hat V$ is the electronic coupling. Expanding the ultrafast excitation in the diabatic basis
\begin{equation}
\hat\mu\ket{E_0} \propto \sum_k \tau_k \ket{d_k}.
\end{equation}
Since this excitation corresponds to an $S_0$ to $S_1$ electronic transition, $\hat P_{S_0} \ket{d_k} = 0$ for all $\tau_k\neq 0$. 
We want to add a correction to each of these $\ket{d_k}$ that only includes components in the $S_1$ electronic state while still being related to the energy eigenstates composing $\hat\mu\ket{E_0}$. Both these requirements are fulfilled if we add a first-order correction to the diabatic wavefunctions using time-independent perturbation theory: using the electronic coupling $\hat{\mathcal{V}}$ as a perturbation, the first-order correction for the diabatic states is
\begin{equation}
\ket{d_k^{(1)}} = \sum_{j\neq k} \frac{\mathcal{V}_{jk}}{E_k^{(0)} - E_j^{(0)}}\ket{d_j}.
\end{equation}
From this, we obtain a first-order corrected basis:
\begin{equation}
\ket{b_k} = \frac{\ket{d_k} + \ket{d_k^{(1)}}}{(\bra{d_k} + \bra{d_k^{(1)}})(\ket{d_k} + \ket{d_k^{(1)}})}.
\end{equation}
The corrected seed vector becomes
\begin{equation} \label{eq:corr_mu}
\ket{\mu^{(1)}} \propto \sum_k \tau_k \ket{b_k},
\end{equation}
where a proportionality constant enforces a norm of unity since the basis $\lbrace\ket{b_k}\rbrace$ is in general not orthonormal.

\section{Scaling analysis} \label{app:scaling}
\subsection{Discrete case}
We now provide a brief analysis on the scaling of the methods with respect to the system size. All methods introduced above consist of an inner iteration (i.e. a single step), and an outer iteration (total number of steps). We analyze both below. \\

For \textit{the outer iterations}, the convergence to the solution will be given by Eqs.\eqref{eq:dyn_explicit}, \eqref{eq:lindblad_gaus}, and \eqref{eq:kraus_ritz} respectively. Defining first the system size $N=N(E_0,E_{max})$ as the number of basis functions necessary for representing the Hamiltonian on some fixed energy interval $(E_0,E_{max})$. We fix the energy interval since otherwise one can always define a larger matrix for a given system which includes unused higher excited states. For a fixed energy interval, the spacing between energy levels will be inversely proportional to the system size (i.e. $\omega_{kj}\sim\mathcal{O}(N^{-1})$).

The convergence of (a) the dynamic averaging approach depends on $\textrm{sinc}(\omega_{kj}t)$, meaning the necessary propagation time, and thus the necessary number of steps, increases linearly with system size $s_{dyn}\sim\mathcal{O}(N)$. For (b) the Lindbladian approach, the convergence depends on $e^{-\omega_{kj}^2 t}$, meaning $s_{lind}\sim\mathcal{O}(N^2)$. For (c) the Lanczos approach, the convergence depends linearly on the number of eigenstates that have a significant contribution to the ultrafast excitation (i.e. bright states). Even though we would typically expect a sublinear growth of the number of bright states with respect to $N$, the exact dependence will depend on the system and the Franck-Condon overlaps. In order to maintain generality, since the number of bright states is bounded by $N$, we recover a convergence that is at most linear with respect to the system size $s_{lanc}\lesssim\mathcal{O}(N)$.  \\

For \textit{the inner iterations}, the Lindbladian method has the worst scaling as well: it requires the propagation of a density matrix, as opposed to the other approaches that only require wavefunction manipulations. Even though the cost of the inner iterations will greatly depend on the method used to solve them, overall a dynamic averaging step is comparable to applying a matrix exponential on a vector ($e^{-i\hat H \Delta t}$), while for the Lanczos approach we require the matrix inverse, which is generally cheaper to implement. We thus expect the Lanczos approach to have a more favorable scaling as the size of the system grows, closely followed by the dynamic averaging approach. Still, the dynamic averaging approach can be done on-the-fly, skipping the need to express the system's Hamiltonian and making the inner iteration cost hard to compare.

\subsection{Continuum case}
Consider now the continuum case, which corresponds to $N\rightarrow\infty$. Since for this case the dimension of the Hamiltonian goes to infinity, a continuum discretization scheme as in Ref.~\citenum{discretization} would be necessary to computationally represent the Hamiltonian and apply the Lanczos method. Still, the dynamic averaging procedure can be applied as long as the time evolution of the wavefunction can be calculated. Its convergence will again be given by Eq.\eqref{eq:dyn_explicit}. Since $\lvert\textrm{sinc}(x)\rvert\leq\lvert x^{-1}\rvert\  \forall x>0$, we will consider a coherence converged when  $x\geq 10$. This corresponds to $\omega_{kj}t\geq 20$. Thus, a propagation for $4.35\times10^6\ \textrm{a.u.} \approx 106 \ \textrm{ps}$ is required for removing coherences between states separated by more than $1 \ \textrm{cm}^{-1}$, while a propagation for $106 \ \textrm{fs}$ will remove coherences between states separated by more than $1000 \ \textrm{cm}^{-1}$.
\bibliography{main}{}

\end{document}